\documentclass[traditabstract]{aa}
\usepackage{graphicx}
\usepackage{txfonts}
\begin{document}
   \titlerunning{Multiple-beam CLEAN for IDVs}
   \authorrunning{I. M. Stewart et al.}
   \title{A multiple-beam CLEAN for imaging intra-day variable radio sources}

   \author{I. M. Stewart \inst{1,2},
           D. M. Fenech \inst{3}
          \and
           T. W. B. Muxlow\inst{1}
          }

   \offprints{I. M. Stewart}

   \institute{Jodrell Bank Centre for Astrophysics, University of Manchester,
             Oxford Road, Manchester M13 9PL, United Kingdom.\\
             \email{Ian.Stewart-2@manchester.ac.uk}
         \and
             Astrophysics, Cosmology and Gravity Centre,
             Department of Astronomy, University of Cape Town,
              Private Bag X3, Rondebosch 7701, South Africa.
         \and
             Department of Physics and Astronomy, University College London,
             Gower Street, London WC1E 6BT, United Kingdom\\
             }

   \date{Received January 0, 0000; accepted January 0, 0000}

   \abstract{
   The CLEAN algorithm, widely used in radio interferometry for the deconvolution of radio images, performs well only if the raw radio image (dirty image) is, to good approximation, a simple convolution between the instrumental point-spread function (dirty beam) and the true distribution of emission across the sky. An important case in which this approximation breaks down is during frequency synthesis if the observing bandwidth is wide enough for variations in the spectrum of the sky to become significant. The convolution assumption also breaks down, in any situation but snapshot observations, if sources in the field vary significantly in flux density over the duration of the observation. Such time-variation can even be instrumental in nature, for example due to jitter or rotation of the primary beam pattern on the sky during an observation. An algorithm already exists for dealing with the spectral variation encountered in wide-band frequency synthesis interferometry. This algorithm is an extension of CLEAN in which, at each iteration, a set of $N$ `dirty beams' are fitted and subtracted in parallel, instead of just a single dirty beam as in standard CLEAN. In the wide-band algorithm the beams are obtained by expanding a nominal source spectrum in a Taylor series, each term of the series generating one of the beams. In the present paper this algorithm is extended to images which contain sources which vary over both frequency and time. Different expansion schemes (or bases) on the time and frequency axes are compared, and issues such as Gibbs ringing and non-orthogonality are discussed. It is shown that practical considerations make it often desirable to orthogonalize the set of beams before commencing the cleaning. This is easily accomplished via a Gram-Schmidt technique.

   \keywords{Methods: data analysis -- Techniques: image processing -- Radio continuum: general}
   }

   \maketitle

\section{Introduction}

It has been known for some decades that some radio sources vary in intensity over time (Dent \cite{dent}). The range of time scales detected so far, roughly from minutes to decades, is set by the practicalities of measurement rather than by any intrinsic properties of the source populations. The present paper is concerned with sources which exhibit significant variation in flux density on time scales of less than 24 hours (so-called Intra-Day Variables or IDVs) and which are also associated with extended structure of a size resolvable by present-day radio interferometers. The structure makes it of interest to image such sources, but the rapid variability can hinder attempts to do so. It is these difficulties in the interferometric imaging of IDV sources which the present paper is designed to address.

How many such sources are there, and of what types? Any estimation of the incidence of variability among the radio source population is complicated by the extra degrees of freedom implicit in a non-flat light curve. It is also inevitable that there will be selection effects due to the inability of the instrument to detect modulation depths below a certain cutoff, or the insensitivity of the observing regime to time scales outside a certain range. The difference between the population of objects observable in our galaxy and those which are extra-galactic introduces a dependence of incidence on galactic latitude (further complicated by the greater propensity of compact, extragalactic sources to scintillate at low galactic latitudes). It is even more difficult to estimate the fraction of variable sources which are IDV, since surveys of the depth and cadence necessary to determine this quantity demand extravagant amounts of observing time.

Of presently-available survey results, De Vries et al (\cite{devries}), in a two-epoch comparison of a high latitude field, found only $\sim 0.2$ deg$^{-2}$ at 1.4 GHz flux densities of $\ge 2$ mJy. Bower et al (\cite{bower}) have processed a large amount of VLA 5 and 8.4 GHz archival data and find a `snapshot' incidence of $1.5$ deg$^{-2}$ radio transients at flux densities greater than 350 $\mu$Jy. Becker et al (\cite{becker}) compared 3 epochs of VLA observations of the galactic plane at 6 cm and found $\sim 1.6$ deg$^{-2}$ sources between 1 and 100 mJy which had a modulation depth greater than 50\%. Regarding the IDV fraction, Kedziora-Chudczer et al (\cite{kedziora}) found, in their multi-frequency study (between 1.4 and 8.6 GHz), that about half of their 13 BL Lac objects, and a smaller fraction of quasars, exhibited IDV up to about 10\% modulation; more recently Lovell et al (\cite{lovell}), in a more comprehensive survey at 5 GHz, report that 37\% of their 443 flat-spectrum sources showed significant variability on a 2-day time scale.

Sources which exhibit a combination of IDV plus some resolved structure fall into several classes, which are briefly reviewed in the following paragraphs.

Novae are expected to exhibit significant changes in radio flux density at intraday time scales. In the standard model, the flux density from an expanding isothermal shell is expected initially to increase proportional to the square of time since the outburst (Hjellming \cite{hjellming}), although some recent VLBA observations appear to be inconsistent with that picture (Krauss et al \cite{krauss}). However, most novae at least in the past decade have not been resolved by radio interferometry until of order 100 days after the outburst, by which time the fractional change in flux density has fallen to about a percent or two per day (see e.g. Eyres et al \cite{eyres_2000}, Eyres et al \cite{eyres_2005}, Heywood \& O'Brien \cite{heywood}). A recent exception is the 2006 outburst of the recurrent nova RS Ophiuchi (O'Brien et al \cite{obrien_2006a}). This was observed with the VLBA first on day 14 after the outburst, and the European VLBI Network (EVN) from day 20 (O'Brien et al \cite{obrien_2006b}). The size of the source at 6 cm wavelength grew from approximately 20 to 35 mas between these observations. The flux density was no longer varying rapidly by that time, but the earliest observations with MERLIN on day 4 after the outburst show it then varying by a factor of 2 over less than a day. Model calculations indicate that RS Oph would have been resolvable by the EVN at that time.

X-ray binaries form an interesting class of galactic IDV sources. Several of these are associated with resolved structure and have emitted jansky-strength radio flares which show pronounced evolution in intensity on intraday time scales. Examples include Circinus X-1 (Haynes et al \cite{haynes}, Tudose et al \cite{tudose_2008}), Cygnus X-3 (Tudose et al \cite{tudose_2009}) and GRS 1915+105 (Fender et al \cite{fender}, Rushton et al \cite{rushton}).

Extragalactic IDV sources fall into fewer categories and are in fact dominated by flat-spectrum AGNs, or objects nowadays collectively described as blazars (Wagner \& Witzel \cite{wagner}, Lovell et al \cite{lovell}). One could almost say that being a blazar is both a necessary and a sufficient condition for an extragalactic source to exhibit IDV, since these objects have the combination of brightness and small size which allows them on the one hand to be detectable at great distances and on the other to be compact enough to exhibit IDV. In many of these objects the compact core appears to be embedded in radio structure which is resolvable by current interferometers (see for example Kovalev et al \cite{kovalev}, Ojha et al \cite{ojha}).

IDV among blazars is an area of current interest. IDV mechanisms for these sources fall into two categories: variation which is intrinsic to the object, and variation caused by the propagation medium (scintillation in the local interstellar medium, or ISS). Both appear to play some role but the relative proportions are not clear. Spectral dependence of the variability at cm wavelengths, annual modulation of the time scales and time shifts between measurements made at different locations all offer strong evidence for the dominance of ISS for some sources (Bignall et al \cite{bignall} and references within). Correlation between variation at radio and other wavelengths (Quirrenbach et al \cite{quirrenbach}) or a bad fit to ISS spectral models (Fuhrmann et al \cite{fuhrmann}) tend to support the intrinsic origin of IDV in others. Intrinsic origin presents theoretical challenges because it is difficult to explain it without invoking either an incidence of extreme Doppler factors which is hard to accept on statistical grounds, or brightness temperatures in excess of the inverse-Compton limit of about $10^{12}$ K (Jauncey et al \cite{jauncey}). Some questions remain also in the ISS picture, such as a relatively weak correlation with galactic latitude (Kedziora-Chudczer et al \cite{kedziora}, Lovell et al \cite{lovell}); the connection between scintillation and scatter-broadening is also still unclear (Ojha et al \cite{ojha}, Lazio et al \cite{lazio}).

The only non-blazar category of extragalactic IDV source which is relevant to the present paper is the maser. McCallum et al (\cite{mccallum}) for example show that the $\mathrm{H}_2 \mathrm{O}$ megamasers in Circinus are both resolvable and exhibit IDV. The techniques described in the present paper can make it easier to image IDV masers which have superimposed lines, meaning they must both be present in the same channel map.

It can thus be seen that there are many objects which it is desirable to image at radio wavelengths with as high resolution as possible, but whose variation on intraday timescales interferes with imaging via earth-rotation aperture synthesis. Broadly speaking, the problem occurs because the IDV causes the pattern of lobes around the time-variable component of the source to be poorly matched to the dirty beam. Such structure cannot be completely removed by the standard CLEAN algorithm, and will therefore tend to limit the dynamic range of the image. As further described in section \ref{clean}, the situation is analogous to the problem in multi-frequency synthesis (MFS), caused by differences in spectral index among objects in the field. The argument for seeking an improved way to image IDV sources is the same as for this MFS case. In fact, as is shown in the present paper, the same technique, originally developed for MFS by Sault and Wieringa (\cite{sault_wieringa}), can be applied to both situations.

The plan of the paper is as follows. Section \ref{interferometry} contains a brief outline of aperture synthesis theory. In sections \ref{conway} and \ref{swa}, the generalized $N$-beam Sault-Wieringa theory is described in detail. The desirability of orthogonalizing the beams is discussed in section \ref{orthogonality}. Different time and frequency basis functions are compared in section \ref{basis_functions}, with emphasis on the avoidance of Gibbs-like phenomena. In section \ref{dual}, a dual expansion in both frequency and time axes is shown to be both necessary and effective in cleaning a simulated wide-band data set which includes sources which vary significantly both in frequency and time. Finally, in section \ref{simpler_lc_method}, a simple 2-beam expansion is derived which is shown to be a powerful means of increasing dynamic range in many IDV cases.

Brief descriptions of the methods described here have already been presented elsewhere in earlier stages of development (Stewart \cite{stewart_2008} and \cite{stewart_2009}).

\section{Interferometry} \label{interferometry}
\subsection{Aperture synthesis} \label{ap_synth}

A radio interferometer measures cross-correlations between the voltage signals detected by pairs of antennas. Each correlation yields a complex number which encodes information about the amplitude and phase of the signals from all sources of radio waves in the field of view of the antennas. Each complex correlation can be considered to be a point sample, plus added noise, of a continuous function $V(\mathbf{u})$, known as the visibility function. The vector $\mathbf{u}$, known as a baseline, is the separation vector between any given pair of antennas, expressed as a number of wavelengths. For an array of antennas which is physically coplanar, $V$ becomes a 2-dimensional function, and can be shown to be the Fourier transform of a function $I(\mathbf{r})$ which is related to the sky brightness distribution at the detector wavelength. Here $\mathbf{r}=(l,m)$ are the sines of the zenith angles of a given sky location. The relation between $I$ and the true sky brightness distribution is given by

\begin{equation} \label{itotrue}
I(\mathbf{r}) = A(\mathbf{r}) \frac{I_\mathrm{true}(\mathbf{r})}{\sqrt{1-\mathbf{r} \cdot \mathbf{r}}},
\end{equation}
\noindent
where $A$, known as the primary beam, is the receptivity of the individual antenna elements as a function of $\mathbf{r}$.

If delays are introduced into the signal chains such that signals from a single point in the sky (known as the phase centre) arrive at the correlator exactly in phase, then the Fourier relationship to the sky brightness distribution becomes approximately true even for non-coplanar arrays, provided the portion of sky to be imaged is restricted to a sufficiently small region around the phase centre. In this general, non-coplanar case, the baseline vectors $\mathbf{u}$ are considered to be the separation vectors between the pairs of antennas projected onto a plane normal to the phase centre. The direction cosines $l$ and $m$ are also in this case taken in a basis in which $(l,m)=(0,0)$ lies in the direction of the phase centre, rather than the zenith.

If the sampling function (which is, to first approximation, a sum of delta functions) is denoted by $S$ then the output of the interferometer is $VS$. The Fourier inversion $\mathfrak{F}^{-1}(VS)$ gives $D=I \ast B$ where $D$, known as the dirty image, is a convolution of the true sky image $I$ with the `dirty beam' $B=\mathfrak{F}^{-1}(S)$. $B$ plays the same role for an interferometer as the point-spread function of a traditional telescope.

In general, the larger the number of samples of $V$, the more compact the distribution of flux density in $B$, and therefore the closer the correspondence between $D$ and the true sky $I$; the sensitivity of the observation will also be increased. An interferometer containing $N$ antennas will generate $N(N - 1)/2$ independent samples of $V$ at each observation. The number of samples can be further expanded via the technique of Earth-rotation synthesis. In this, the rotation of the Earth over the course of a day is used to generate a sequence of different baselines $\mathbf{u}$ for each antenna pair. Another way to increase the number of samples is to observe at several frequencies, a technique known as frequency synthesis. Because baselines are expressed in wavelengths, the fixed spatial separation between a pair of antennas generates baseline vectors $\mathbf{u}$ of different length at different frequencies. Implicit in these two synthesis techniques is the assumption that the sky brightness will be constant over time in the first case and over frequency in the second.

A more detailed description of the fundamentals of interferometry can be found in many sources, for example Thompson, Moran \& Swenson (\cite{thompson}) or Taylor, Carilli \& Perley (\cite{syn_im_2}).

\subsection{CLEAN} \label{clean}

The CLEAN algorithm was invented by H\"{o}gbom (\cite{hoegbom}) and has been further elaborated by Clark (\cite{clark}) and Cotton and Schwab (Schwab \cite{schwab}) among others. The essence of the algorithm is to perform many iterations of a process in which a small amount of the dirty beam is subtracted, centred at the highest remaining point in the dirty image, ideally until nothing remains in this image but noise. The positions and amounts subtracted are recorded as `clean components' and used afterwards to reconstruct an approximation to the true sky image $I$. The gain factor by which the dirty beam is multiplied, and the number of iterations to perform, are parameters which are chosen ahead of time by the user.

CLEAN was originally presented as an empirical algorithm which appeared to produce results, although it required some experience to judge how best to apply it. It is known not to perform well when applied to extended objects (Cornwell \cite{cornwell_1983}), although a modified algorithm has been shown to yield improvements here (Steer, Dewdney \& Itoh \cite{steer}). CLEAN is still in wide use as a practical method of removing sampling artifacts from interferometry images.

The reasons for CLEAN's success were unclear for some time (Cornwell et al \cite{cornwell_1999}), and its theoretical basis has come under occasional criticism (Tan \cite{tan}, Lannes et al \cite{lannes}). Only relatively recently has it been shown to be related to compressive sampling (Cand\`{e}s \& Wakin \cite{compressive_sampling}) and been given a sound theoretical underpinning.

In order for H\"{o}gbom CLEAN to work, the convolution relation $D=I \ast B$ must be valid. There are cases where this assumption breaks down however. One of the most severe departures occurs in frequency synthesis with large fractional bandwidths. It is common for a single observational field to contain objects whose spectral indices differ by several tens of percent or more. Differences in spectra of this order are unimportant if the observation fractional bandwidth is small but may significantly degrade the convolution assumption where the fractional bandwidth approaches 1.

Conway et al (\cite{conway}) showed that, in the wide-band case in which the dirty image $D$ no longer well approximates a convolution of the sky brightness distribution $I$, it was nevertheless possible to express $D$ as a sum over a relatively small number $N$ of component images $D_i$, each of which individually obeys a convolution relation $D_i=I_i \ast B_i$. Conway et al arrived at this by expanding the nominal spectrum at each sky location in a Taylor series, each $i$th term of the series generating a respective `spectral dirty beam' $B_i$. In this case each image $I_i$ is simply a sky map of the value of the $i$th Taylor coefficient. Conway et al suggested a coordinate transform for better application of this technique to commonly-found power-law radio spectra, and presented an approximate method to solve for the coefficient images $I_i$ when $N$ is restricted to 2. This method consists of a 2-step sequential CLEAN and relies on the beams $B_0$ and $B_1$ being approximately orthogonal.

Sault and Wieringa (\cite{sault_wieringa}) retained the Taylor expansion but devised a new solution algorithm which can be thought of as a generalization of the H\"{o}gbom CLEAN algorithm from its original `scalar' context, in which a single image $D$ is iteratively deconvolved, to a new `vector' context in which $N$ images $D_i$ are deconvolved in parallel. Orthogonality of the beams is no longer required (although the technique fails if 2 or more beams are identical). Sault and Wieringa elaborated their theory as it applied to the $N=2$ case but, as the authors themselves suggest, the extension to $N>2$ is not difficult.

The CLEAN algorithm continues to be a subject of active development. Recent work includes an extension of CLEAN to produce clean components with a range of sizes (Cornwell \cite{cornwell_2008}), a modification to clean tomographic LISA images (Hayama et al \cite{hayama}), and an adaption of the algorithm to reconstruct RHESSI images of solar flares (Schwartz \cite{schwartz}). The compressive-sampling formalism has recently also generated some promising new approaches (Wiaux et al \cite{wiaux}; Li, Cornwell \& de Hoog \cite{li}).

\section{A generic multiple-beam CLEAN} \label{generic_clean}
\subsection{Generalized Conway decomposition} \label{conway}

In its most general form, the aim of the treatment described by Conway et al (\cite{conway}) is to find a way to decompose the dirty image into a sum of convolutions
\begin{equation} \label{equ_a}
  D \sim \sum_{i=0}^{N-1} D_i = \sum_{i=0}^{N-1} I_i \ast B_i.
\end{equation}
The problem is to calculate a set of beams $B_i$ for which this relation is true. There are no doubt many ways to do this, but in the present section we are going to consider only the route which comes from expanding the frequency and time dependence of the sky brightness in a sum of basis functions. Such an expansion is written
\begin{equation} \label{equ_b}
  I(\mathbf{r},\nu,t) \sim \sum_{p=0}^{P-1} \sum_{q=0}^{Q-1} I_{p,q}(\mathbf{r}) \, F_p(\nu) \, T_q(t),
\end{equation}
where $I(\mathbf{r},\nu,t)$ is the brightness function; $\mathbf{r}=(l,m)$ are direction cosines of the angular displacement from the phase centre; $\nu$ is frequency and $t$ time; $I_{p,q}(\mathbf{r})$ is a sky map of the $(p,q)$th component; and $F_p$ and $T_q$ are $p$th and $q$th members respectively of sets of basis functions. Earth-rotation and frequency synthesis will return a set of $j=1$ to $M$ measurements $V_j$ of the cross-correlations between antennas. Ignoring for the sake of simplicity non-coplanarity, weights, the shape of the primary beam, and the spherical projection correction $(1-\mathbf{r}\cdot\mathbf{r})^{-0.5}$ (see equation \ref{itotrue}), each $V_j$ is related to $I$ by a transform:
\begin{displaymath}
  V_j = \int \! d\mathbf{r} \, \exp(-2\pi i \, \mathbf{u}\cdot\mathbf{r}) \, I(\mathbf{r},\nu_j,t_j).
\end{displaymath}

Although each $V_j$ represents an average over a small time $\times$ bandwidth window, if this window is small enough, $V_j$ may be represented as a point sample of the visibility function at the spatial frequency $\mathbf{u}_j$. The total set $\mathbf{V}$ of visibility measurements can thus be written
\begin{displaymath}
  \mathbf{V} = \sum_{j=1}^M \delta(\mathbf{u}-\mathbf{u}_j,\nu-\nu_j,t-t_j)  \int \! d\mathbf{r} \, \exp(-2\pi i \, \mathbf{u}\cdot\mathbf{r}) \, I(\mathbf{r},\nu,t).
\end{displaymath}
Expanding $I$ in its basis functions gives, with some rearrangement:
\begin{displaymath}
  \mathbf{V} = \sum_{p,q} \sum_{j=1}^M \delta(\mathbf{u}-\mathbf{u}_j) \, F_p(\nu_j) \, T_q(t_j) \int \! d\mathbf{r} \, \exp(-2\pi i \, \mathbf{u}\cdot\mathbf{r}) \, I_{p,q}(\mathbf{r}).
\end{displaymath}
Fourier inversion of $\mathbf{V}$ yields the dirty image $D$. This is indeed found to be a sum of convolutions:
\begin{equation} \label{equ_f}
  D = \mathfrak{F}^{-1}(\mathbf{V}) = \sum_{p,q} I_{p,q} \ast B_{pq},
\end{equation}
where
\begin{displaymath}
  B_{pq} = \mathfrak{F}^{-1} \left( \sum_{j=1}^M \delta(\mathbf{u}-\mathbf{u}_j) \, F_p(\nu_j) \, T_q(t_j) \right).
\end{displaymath}
Here the $\mathfrak{F}$ symbol represents the Fourier transform. The two sums over $p$ and $q$ are easily collapsed to one, in which case equation \ref{equ_f} maps directly to equation \ref{equ_a}. Clearly this requires that $N=P \times Q$.

\subsection{The $N$-beam Sault-Wieringa algorithm} \label{swa}

Sault and Wieringa (\cite{sault_wieringa}) developed an algorithm which, although it was set down only for $N=2$, is easily generalized to the following:
\begin{itemize}
  \item Generate initial $R_i$ from $D \star B_i$ (the $\star$ here represents correlation).
  \item Generate, for $i \in [0:N-1]$ and $j \in [0:N-1]$, cross-correlation images $Z_{i,j} = B_i \star B_j$.
  \item Calculate a matrix $\mathbf{M}$ such that its $(i,j)$th element is given by $Z_{i,j}(\mathbf{r}=0)$.
  \item Perform a number of iterations of the following procedure:

  \begin{enumerate}
    \item Find the location $\mathbf{r}_\mathrm{max}$ of the maximum value of the image formed from $R = \sum_{i,j} R_i R_j M^{-1}_{i,j}$.
    \item Perform, for each $i$, the subtractions
\begin{displaymath}
  R_i(\mathbf{r}) = R_i(\mathbf{r}) - \lambda \sum_{j=0}^{N-1} R_j(\mathbf{r}_\mathrm{max}) Z_{i,j}(\mathbf{r}-\mathbf{r}_\mathrm{max}).
\end{displaymath}
    \item Form a vector $\mathbf{c}_\mathrm{raw}$ of the entangled clean components $\lambda R_i(\mathbf{r}_\mathrm{max})$.
    \item Solve the equation $\mathbf{c} = \mathbf{Mc}_\mathrm{raw}$ to disentangle the clean components.
    \item Store these for later cleaned-image reconstruction.
  \end{enumerate}
\end{itemize}

\subsection{Orthogonality pros and cons} \label{orthogonality}

The Sault-Wieringa deconvolution includes a matrix inversion. If two or more beams from the set of $N$ are identical, the matrix $\mathbf{M}$ is singular, i.e. uninvertible. If no two beams are exactly identical, but two or more are similar (have a normalized scalar product close to unity), $\mathbf{M}$ will be formally invertible but can be expected to be ill-conditioned (i.e., result in large errors in the final clean components). Hence approximate orthogonality between beams is desirable; but if this be established, little further reduction in error is to be gained by enforcing complete orthogonality.

There may however be other reasons to manipulate the beams, even to the point of transforming them into an orthonormal set, other than keeping the matrix $\mathbf{M}$ well-conditioned. Two such circumstances are described here. Firstly, the main aim of the Conway decomposition and subsequent vector cleaning is arguably to obtain a better image of the average sky brightness distribution. This image is identical to the image of the zero-order beam coefficient if and only if every other beam is zero-valued at its centre - because a non-zero value of $B_i(0)$ equates to a non-zero average of the $i$th basis function over the frequency and time range of the observation. Thus in order to simplify the post-production of an average-brightness image, one would want to modify the set of beams before cleaning by subtracting $B_i(0) \times B_0 / B_0(0)$ from each $B_i$ for $i > 0$.

The second circumstance concerns the practicalities of computing the deconvolution. The Sault-Wieringa algorithm requires that $N(N+1)$ images be kept in memory during the iteration. For $N=2$, as treated by Sault and Wieringa, this is a manageable number of images. However, some of the scenarios discussed in the present paper make use of values of $N$ as high as 30. Maintenance within memory of close to 1000 arrays, each of which may be as large as 2048 x 2048 pixels, is likely to prove a strain for most machines at least in the lower echelons of present-day computing power. The load may be much reduced by orthogonalizing the beams before the start of the algorithm, i.e. by generating a set of modified beams $B_i'$ such that $B_i' \cdot B_j' = 0$ for all $i \not= j$. The cross-correlation images $Z_{i,j}'$ generated from these will be such that their central values, thus the elements of $\mathbf{M}^\prime$, will be 0 for all $i \not= j$. There is no need to store or even generate the `off-diagonal' images; the necessary inventory of images thus reduces to $2N$.

The transformation from the basis functions in equation \ref{equ_b} to beams in `sky-image' space does not preserve orthogonality. Because of this there seems little to be gained by requiring the basis functions to be orthogonal: one must work directly with the dirty beams themselves.

The Gram-Schmidt process of converting $B_i$ to orthonormal $B_i'$ can be expressed as the matrix equation
\begin{equation} \label{equ_h}
  \mathbf{b} = \mathbf{G} \mathbf{b}',
\end{equation}
where $\mathbf{b}$ is a vector of the $N$ input dirty beams $B_i$, $\mathbf{b}'$ is a vector of the orthonormalized beams $B_i'$, and $\mathbf{G}$ is a lower-triangular matrix with elements defined as follows: $G_{i,i} = B_i' \cdot B_i'$ before $B_i'$ is normalized, and $G_{i,j}$ for $j < i = B_j' \cdot B_i$.

As Golub \& Van Loan (\cite{gandvl}) point out, the usual Gram-Schmidt procedure is known to be numerically unstable. These authors suggest an alternate, stabilized algorithm which amounts to solving equation \ref{equ_h} by backward substitution, the elements of $\mathbf{G}$ being calculated on the fly.

The Sault-Wieringa algorithm, working with orthonormalized beams, produces, for each image pixel $(x,y)$, a vector of clean components $\mathbf{c}'$ such that
\begin{displaymath}
  D = \sum c_i' B_i',
\end{displaymath}
or, in vector notation,
\begin{equation} \label{equ_j}
  D = \mathbf{c}^{\prime \mathrm{T}} \mathbf{b}'.
\end{equation}
How to reconstruct the clean components which give $D$ in terms of the original beams? Inverting equation \ref{equ_h} and inserting \ref{equ_j} gives
\begin{displaymath}
  D = \mathbf{c}^{\prime \mathrm{T}} \mathbf{G}^{-1} \mathbf{b}.
\end{displaymath}
Obviously the desired clean components $\mathbf{c}$ are given by
\begin{displaymath}
  \mathbf{c}^\mathrm{T} = \mathbf{c}^{\prime \mathrm{T}} \mathbf{G}^{-1}.
\end{displaymath}

\subsection{Post-processing} \label{reconstruction}

The output of the Sault-Wieringa process is, as with H\"{o}gbom CLEAN, a list of clean components; but here each component is no longer scalar but is itself an $N$-element vector or list. The question then arises of what to do with this information. Almost certainly one will want to make an image of the average sky brightness, since one of the motivations of the Conway decomposition has been to obtain a more accurate reconstruction of the sky, free from the spectral artifacts which one sees in H\"{o}gbom-based deconvolution of wide-band observations. One might think that similar images should also be made from each of the higher-order elements of the clean components. Experience shows however that such images rapidly become too noisy to be of use as the order increases. In broad terms this is because the higher-order basis functions tend to vary more rapidly with displacement in the UV plane; which means they contain more power in higher spatial frequencies in this domain; which translates to a broader, less centrally peaked structure for the respective beam in the sky domain. This broadening and diluting of the beam profile for higher orders (visible in the example beam profiles of Conway et al \cite{conway}) causes relatively more `false detections' at these orders to occur during the deconvolution, until at the end the real clean component values for this order may become lost among the false ones. The problem is exacerbated by the tendency of the true component values to decrease (with any sensible choice of basis function) as the order increases.

A possible remedy for this problem might be to choose smaller values of the loop gain $\lambda$ as the order of the basis function increases. But we have not so far tested this conjecture.

For constructing a restored image of the average brightness, it seems clear that the best thing to do is construct an average-brightness beam, and fit the restoring beam to its centre, in analogy to the present standard practice when employing H\"{o}gbom CLEAN. However, there is no such obvious source of a restoring beam profile for higher-order images, since their respective beams don't necessarily have a peak in the centre to fit to. Use of the average-derived restoring beam in these cases is probably acceptable though, since the underlying spatial resolution should be similar. Another possibility would be to use a Gaussian of the same full width half maximum as the beam.

\section{Time-variable sources} \label{time_var}
\subsection{Expansion in basis functions} \label{basis_functions}

It is a commonplace that radio continuum spectra tend to be smooth. In nearly all cases a continuum spectrum may be well approximated over an octave or so (a frequency range which the forthcoming wide-band interferometers are not expected to greatly exceed) by a polynomial of at most 2 or 3 orders. In this case there seems little to be gained by investigating the use of other types of basis function. Light curves, on the other hand, are much less well behaved: fluctuations can be expected over a wide range of timescales. Sault-Wieringa style parallel cleaning of observations which include time-variable sources may therefore require expansion of the nominal light curve to higher order than is usually necessary for spectra. This also makes it of more interest to compare different basis expansions in this case.

Note however that an observation which includes only a single variable source can be cleaned via the simple 2-beam technique described in section \ref{simpler_lc_method}.

All other aspects being equal, one would choose the basis set which resulted in the smallest residuals for a given $N$ in equation \ref{equ_a}, which is another way of saying that the sum over $N$ `component' dirty images $D_i$ should converge rapidly to the `true' dirty image $D$ as $N$ increases. And rapid convergence in the sky plane is surely linked to rapid convergence of the basis function expansion in the U-V plane. However, it is difficult to predict which set will best meet this criterion, for two reasons. Firstly the light curves encountered in practice may be very diverse: a basis set which converges rapidly to one sort may not do so for another. Secondly, there appears to be at present no analysis which predicts the image residuals. Thus trial and error, as with traditional CLEAN, probably remains the best guide here too.

In the present section we compare two sets of basis functions for the expansion of light curves, namely Fourier sinusoids and Chebyshev polynomials.

If a standard Fourier series is employed, the expansion will suffer from Gibbs phenomenon at the boundaries, unless the light curve is periodic continuous (in fact periodic analytic) over the chosen observation interval. Gibbs ringing can however be much reduced by the following treatment. Let the duration of the observation be denoted by $T$, and the light curve at a given sky direction by $f(t)$. Suppose one doubled the observation time, and filled the new interval with a function $g$ defined such that
  \[ g(t) = \left\{ \begin{array}{l}
     f(t), 0 \le t < T\\
     f(T-t), T \le t < 2T. \end{array} \right. \]
The new function $g$ is periodic continuous in the interval $[0,2T]$, so there should be no zeroth-order Gibbs ringing in its Fourier expansion. $g$ is also symmetric about zero, so only cosine terms will remain in the expansion. If the resulting Fourier basis functions are truncated back to $[0,T]$ they are seen to be given by
\begin{equation} \label{equ_m}
  T_n(t) = \cos(\pi n t/T) \mbox{ for } n \in [0,..N].
\end{equation}
Similar tricks can be used to enforce boundary differentiability to higher orders if desired.

If there are gaps in the time sequence of data values - patches of bad data perhaps, or even planned, periodic observations of a phase calibrator - such gaps might be expected to cause additional Gibbs-type problems. A normal Fourier expansion of a function which suddenly drops to zero at intervals would suffer Gibbs ringing at the discontinuities. However, Conway decomposition is not `normal expansion' in the basis functions. The actual basis onto which the dirty image is projected consists of the set of beams; and the construction of each beam from the corresponding basis function in the UV plane effectively disconnects it from any irregularities in the support of the basis functions. Orthonormal basis functions won't necessarily produce orthonormal beams, and vice versa.

A final point to note is that this technique cannot disentangle or unwrap an observation which spans more than 24 hours at the same frequency: in this case all that can be obtained is the net (cyclic superimposed) light curve.

\subsection{Simulations} \label{simulations}
\subsubsection{A single time-variable source} \label{single-source}

\begin{figure}
   \centering
      \resizebox{\hsize}{!}{\includegraphics[angle=-90]{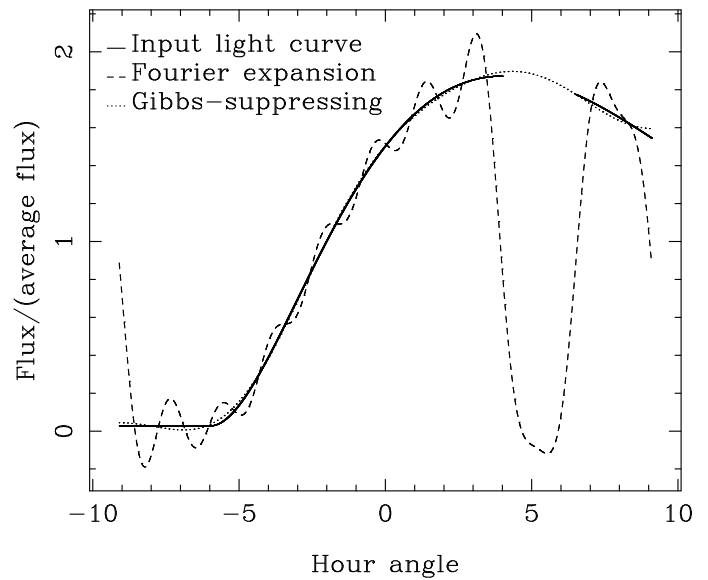}}
      \caption{Direct Fourier expansion of a simulated light curve compared to the result of a Sault-Wieringa clean using a set of basis functions which suppresses Gibbs oscillation at the boundaries.
              }
         \label{fig_a}
   \end{figure}

\begin{figure}
   \centering
      \resizebox{\hsize}{!}{\includegraphics[angle=-90]{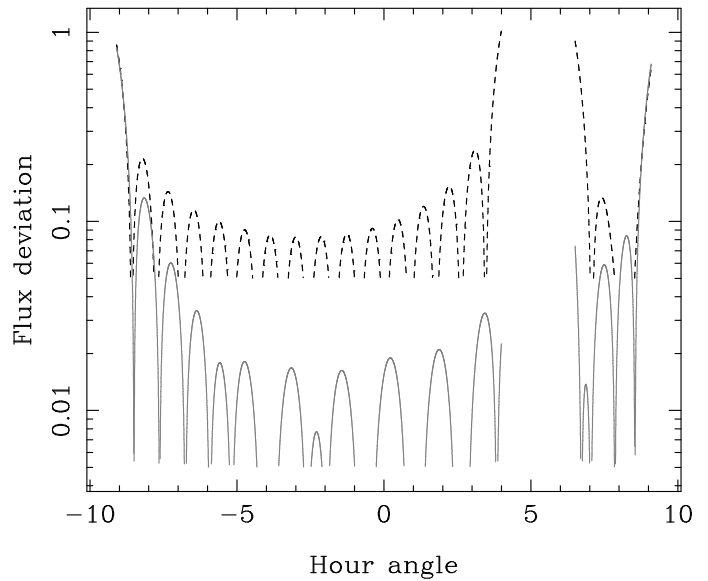}}
      \caption{The same light curve as in figure \ref{fig_a} but now with residuals plotted. Shown are the residuals from the Fourier expansion of figure \ref{fig_a} (dashed curve) and the residuals from a Sault-Wieringa clean using the same set of Fourier basis functions (half-tone solid curve). For better clarity, values of small magnitude have not been plotted; and a different cutoff was used for each curve. Points falling within the data gap have also been omitted.
              }
         \label{fig_b1}
   \end{figure}

\begin{figure}
   \centering
      \resizebox{\hsize}{!}{\includegraphics[angle=-90]{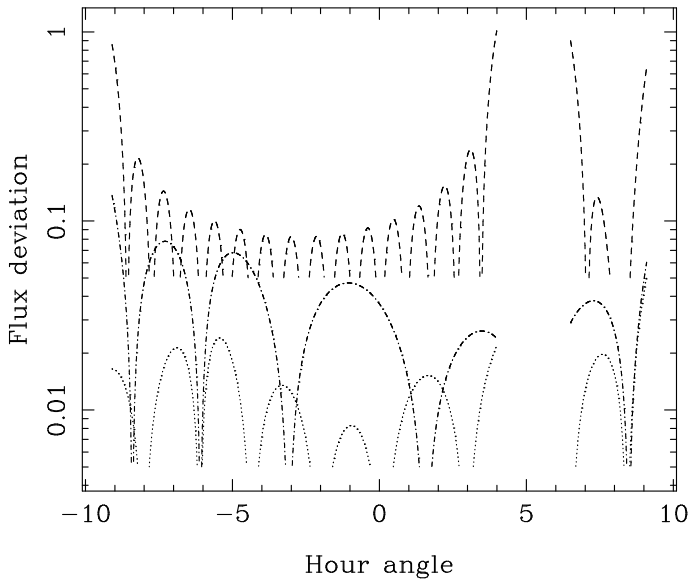}}
      \caption{This is similar to figure \ref{fig_b1}. The dashed curve again represents the Fourier residuals, whereas the dash-dot and dotted curves are residuals from a Sault-Wieringa clean using two different types of basis function. For the dash-dot curve, Chebyshev polynomials up to order 5 were employed to generate the Sault-Wieringa beams; the dotted curve is the result of using the half-frequency cosine functions (also to the order 5).
              }
         \label{fig_b2}
   \end{figure}

\begin{figure}
   \centering
      \resizebox{\hsize}{!}{\includegraphics[angle=-90]{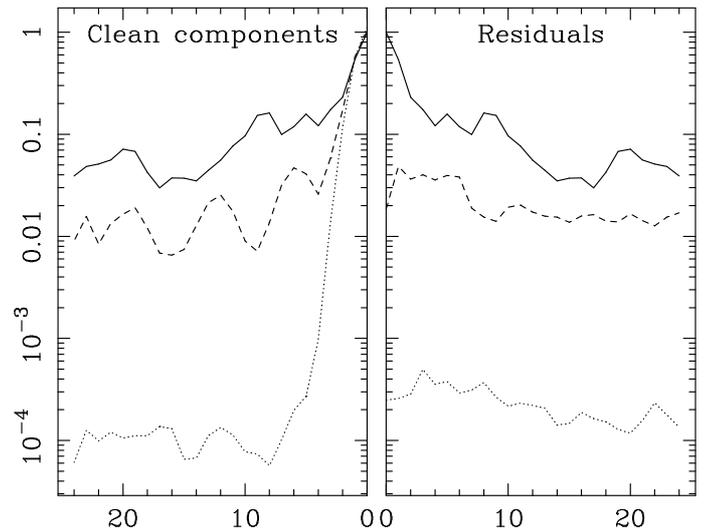}}
      \caption{Both sides of this plot show RMS image values as a function of radius from the source position. Values calculated from clean-component images appear in the left panel; those on the right come from residual images. The radial distribution of RMS values in the dirty image (solid line) is included in both panels for purposes of comparison. The dashed lines represent the H\"{o}gbom-cleaned data, whereas the dotted lines are from the Sault-Wieringa cleaning.
              }
         \label{fig_c}
   \end{figure}

Figures \ref{fig_a} to \ref{fig_c} refer to a simulation constructed as follows. Artificial visibilities were calculated for a single time-varying source which was located at the phase centre. The array parameters were those of MERLIN; the chosen source declination was $+40^\circ$; integration time was 10s, and 32 channels of 1 MHz width starting at 6 GHz were used. Perfect calibration was assumed and no instrumental noise was added. A gap was introduced into the data sequence between approximately 4 and 6.5 hours postmeridian.

Figures \ref{fig_a}, \ref{fig_b1} and \ref{fig_b2} show the effect of different basis expansions on the Gibbs phenomenon; figure \ref{fig_c} concerns the relative efficiency of H\"{o}gbom versus Sault-Wieringa cleans for time-variable sources.

In figure \ref{fig_a}, the light curve of the source (solid line) is compared firstly to a direct Fourier expansion of this light curve to order 10 (dashed line), and secondly to the light curve reconstructed after Conway decomposition into 6 beams (i.e. to order 5) from the half-frequency cosine basis functions prescribed by equation \ref{equ_m}, followed by 1000 cycles of Sault-Wieringa cleaning with loop gain 0.1 (dotted line). The Gibbs ringing of the standard Fourier expansion at boundaries and other discontinuities is very obvious, as is the almost complete absence of such an effect in the Sault-Wieringa result.

Figure \ref{fig_a} shows clearly that Gibbs phenomenon can be largely avoided in Sault-Wieringa cleaning of time-varying sources. It is of interest however to explore a little more deeply into the difference between boundary discontinuities and data gaps, and the relative efficacy of different basis functions. Figures \ref{fig_b1} and \ref{fig_b2} serve this purpose. For these figures, the same input data are used, but the plotting style is different to figure \ref{fig_a}. Here are plotted, on a logarithmic scale, absolute values of the residuals of the curves of interest against the input light curve.

In figure \ref{fig_b1} the direct Fourier expansion (dashed line) is compared to the result of a Sault-Wieringa clean (half-tone solid line) as in figure \ref{fig_a}; but here the unmodified Fourier functions, i.e. the same functions used in the direct expansion, were chosen as the basis set (which generated 21 beams!). As expected, in both cases there is Gibbs oscillation at the boundary, but only in the direct expansion is there also Gibbs at the edges of the data gap. The conclusion here is that it is unnecessary to choose the basis set with any care in order to avoid Gibbs ringing at data gaps: the Sault-Wieringa technique by its nature avoids such troubles. The same is however obviously not true of the problem at the boundaries, which arises from the inability of the Fourier basis set to represent a function with a discontinuity at the boundary. In effect, the Sault-Wieringa deconvolution can interpolate over a gap, but cannot (without `outside help') deal with a discontinuity.

Figure \ref{fig_b2} again shows residuals from the direct Fourier expansion (dashed line) but compares them here to residuals from two different Sault-Wieringa deconvolutions: using firstly the half-frequency cosine basis of equation \ref{equ_m} (dotted line, as already displayed in figure \ref{fig_a}), and secondly, Chebyshev polynomials (dot-dashed line). Both expansions were truncated at order 5. It is perhaps slightly unexpected that the cosine functions seem to fit the light curve better than the Chebyshevs.

Figure \ref{fig_c} compares a H\"{o}gbom clean (1000 cycles at gain 0.1) of the same time-varying single source with Sault-Wieringa clean using the half-frequency cosine basis to order 5. (The Sault-Wieringa result is the same which generated the dotted curves in figures \ref{fig_a} and \ref{fig_b2}.) The data plotted in figure \ref{fig_c} come from images: the dirty image for the data set, and the clean and residual images from the H\"{o}gbom and Sault-Wieringa cleans respectively. What has been done for each image is to calculate the RMS of the image values in a polar coordinate system centred on the source at the phase centre. The RMS values have been plotted as a function of radius from this centre.

Use of the Sault-Wieringa procedure is seen to improve the dynamic range by about 2 orders of magnitude to almost $10^{4}$.

\subsubsection{Several sources varying in frequency and time} \label{dual}

\begin{figure}
   \centering
      \resizebox{\hsize}{!}{\includegraphics[angle=-90]{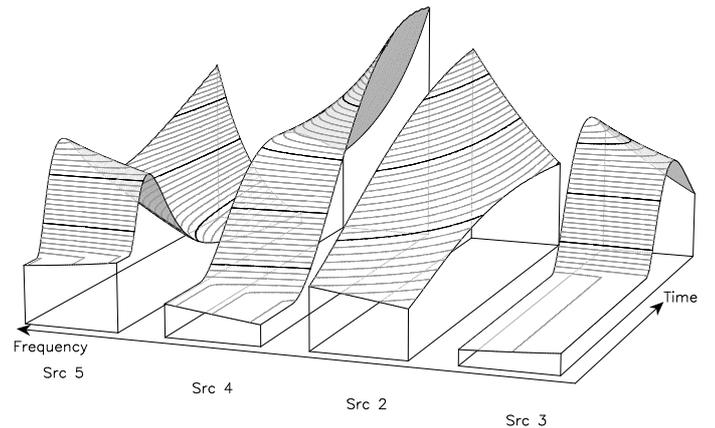}}
      \caption{Surfaces showing the frequency and time dependence of the 4 variable sources. The height of each surface represents the flux density of that source. Major contours occur at 1 jansky intervals. The non-variable source had a flat spectrum at 1 Jy. Note that the order of the sources has been slightly rearranged for better display of the surfaces.
              }
         \label{fig_g}
   \end{figure}

The simulation described in the present section was devised to be an exacting test of the Conway/Sault-Wieringa technique as applied to sources which vary both in frequency and time. The input data were derived from a model containing five point sources, each of which had a power-law spectrum. The first source was to serve as a reference, and so had a flat spectrum with no time variation. For the remaining sources, both the average flux density and the spectral index were made to vary over time. The frequency and time dependences of the remaining sources are illustrated in figure \ref{fig_g}.

The array parameters for this simulation were again those of MERLIN, with a phase centre declination of $+40^\circ$; however the bandwidth chosen this time spanned from 5 to 7 GHz, divided into 50 channels of width 40 MHz. The integration time was set to 60 sec. These chosen values for channel width and integration time are much coarser than those usually associated with MERLIN, and would set severe limits to the breadth of field in a real observation. In the present case, the field of interest is narrow, so coarse values were chosen in order to reduce the computing load. The chosen values were the largest ones consistent with accurate gridding of the relatively small area of the field occupied by the simulated sources.

The simulated visibilities were gridded with uniform weighting and subjected separately to 2000 cycles of H\"{o}gbom cleaning at a loop gain of 0.1 versus the same number of cycles, at the same gain, of Sault-Wieringa cleaning, using $FT$ basis functions as described in section \ref{conway}, the $F$ functions being Chebyshev polynomials (although to the small order used, these are scarcely distinguishable from Taylor series terms) and the $T$ functions being the half-frequency Fourier cosine functions of equation \ref{equ_m}.

\begin{figure}
   \centering
      \resizebox{\hsize}{!}{\includegraphics[angle=0]{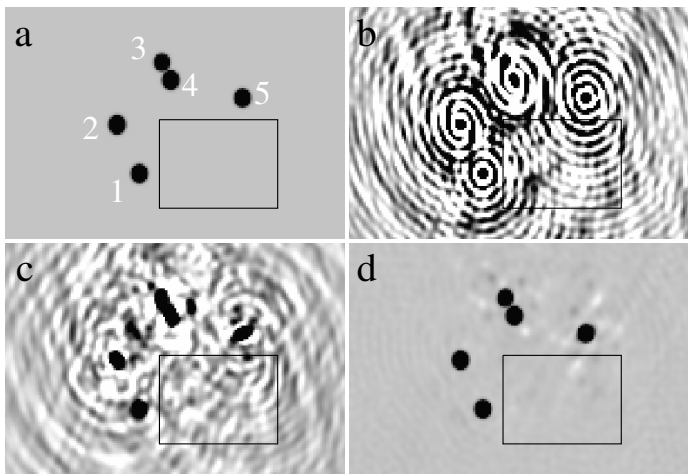}}
      \caption{Figure $\mathbf{a}$ is an image constructed from the average flux densities of the input sources to the simulation. $\mathbf{b}$ is the dirty image. $\mathbf{c}$ is the result of 2000 cycles of H\"{o}gbom clean at a loop gain of 0.1. $\mathbf{d}$ is the result of a Sault-Wieringa clean (with the same niter and loop gain) as described in the text. The unsaturated brightness range for all four images was -0.02 to +0.02 Jy/beam. (For comparison purposes, the brightest pixel of the source image was 1.27 Jy/beam.) The grey scale is reversed. Black rectangles indicate the area chosen for the RMS measurement as described in the text.
              }
         \label{fig_d}
   \end{figure}

A number of images relating to this simulation are shown in figure \ref{fig_d}.  

An RMS value was calculated for those pixels within a rectangular area which is shown on the plots by a black outline. These numbers are shown in table \ref{tab_a}. The Sault-Wieringa technique clearly offers a significant improvement in dynamic range.

   \begin{table}
   \caption{RMS values from cleaned images.}             
   \label{tab_a}      
   \centering                          
   \begin{tabular}{c c}        
   \hline\hline                 
   Image name & RMS (Jy)\\    
   \hline                        
   Dirty image     &  1.588e-2 \\      
   H\"{o}gbom cleaned &  4.847e-3 \\
   SW cleaned      &  5.415e-4 \\
   \hline                                   
   \end{tabular}
   \end{table}

As discussed in section \ref{reconstruction}, the value of this technique lies more in its ability to remove artifacts from an image of average flux density, rather than as a way to estimate light curves or spectra. For wide-band observations in which there is no time variation over the observation, because of the typically slowly-varying nature of radio spectra, one may reasonably expect to extract low-order spectral information (such as spectral indices) from Conway decomposition. But because light curves may easily contain significant power at higher orders of the time basis functions, which are not so accurately recovered by the deconvolution, more caution is advisable in interpreting the time-dependent output of the deconvolution process. If an accurate light curve of a source is desired, it is probably always going to be preferable to just phase-rotate the array to that sky location.

\subsubsection{Simpler method for a single variable source} \label{simpler_lc_method}

A source which shows significant time variation within the course of a day must have a spatial dimension less than about a light-day across. When observed with, for example, the MERLIN array at 21 cm, any such source more distant than about a kiloparsec will be unresolved. Except in the case of masers, it is also unusual to find two such variable sources in close proximity. For many observations of time-variable radio sources, therefore, we may expect the source to be point-like (unresolved) and the only source in the field which shows a significant time variation. In this case the light curve for the whole observed field, $s_\mathrm{total}$, is just a simple sum of the source light curve $s_\mathrm{source}$ plus a time-invariant background flux density. Any point in the image can therefore be expressed as a sum of just two basis functions:
\begin{eqnarray*}
  T_0(t) & = & 1\\
  T_1(t) & = & s_\mathrm{total} - \langle s_\mathrm{total} \rangle.
\end{eqnarray*}
Here the mean notation $\langle \rangle$ is used to indicate that beam 1 is constructed as follows. A raw beam 1 is first formed by gridding, weighting and transforming the visibilities from a source at the phase centre which has a light curve given by $s_\mathrm{total}$. An amount of beam 0 is then subtracted from it such that the central value of the result is zero. Beam $B_1$ is adjusted in this way so that the average flux-density information over the field is contained entirely in the distribution of component 0.

A further simulation was constructed to test this technique. This simulation contained a time-variable point source located at the phase centre together with much fainter, extended emission (actually made up of 24 closely-spaced point sources) extending over about 0.2 arcsec (equal to 20 image pixels) either side of the central source. The average flux density of the central source was 1 Jy/beam whereas the extended emission ranged in brightness from about $3.5\times 10^{-3}$ to $1.5\times 10^{-3}$ Jy/beam. The light curve of the central source was the same as diagrammed in figure \ref{fig_a}, but without the data gap. It brightens by 3.7 magnitudes in the course of the observation.

\begin{figure}
   \centering
      \resizebox{\hsize}{!}{\includegraphics[angle=0]{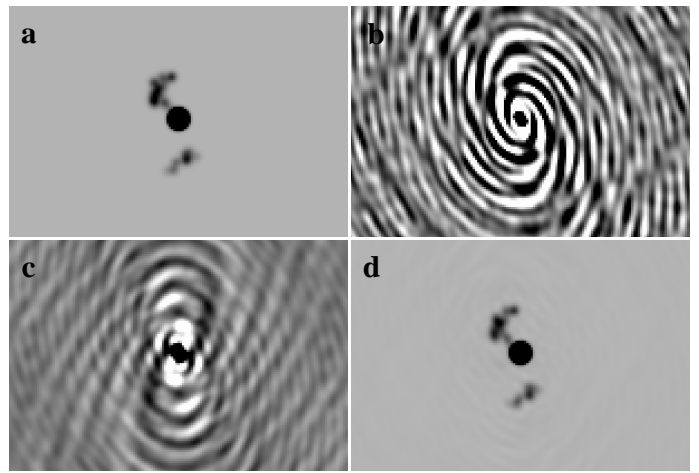}}
      \caption{Figure $\mathbf{a}$ is an image constructed from the average flux densities of the input sources to the simulation. $\mathbf{b}$ is the dirty image. $\mathbf{c}$ is the result of 5000 cycles of H\"{o}gbom clean at a gain of 0.01. $\mathbf{d}$ shows the distribution of the zeroth component of a Sault-Wieringa clean, with similar gain and number of clean cycles, using the two-beam decomposition as described in the text. The unsaturated brightness scale was -0.005 to +0.005 for $\mathbf{a}$ and $\mathbf{d}$, -0.05 to +0.05 for $\mathbf{b}$ and $\mathbf{c}$. The grey scale is reversed for all images.
              }
         \label{fig_e}
   \end{figure}

A quartet of images (similar to figure \ref{fig_d}) to exhibit the performance of this technique is shown in figure \ref{fig_e}. For constructing the visibilities, MERLIN specifications were again used. The integration time was 5 seconds, and 32 channels of width 1 MHz, starting at 6 GHz, were specified.

It is easily seen that the Sault-Wieringa deconvolution recovers almost all the faint emission. The H\"{o}gbom clean is unable to remove sidelobes at a level 10 times the flux density of the extended emission and thus is incapable of revealing it. Several values of the H\"{o}gbom gain and number of iterations were tried with no improvement on what is displayed here.

\section{Conclusions}

In this paper, a technique developed originally by Conway et al (\cite{conway}) and Sault and Wieringa (\cite{sault_wieringa}) to allow cleaning of multi-frequency-synthesis images has been generalized and shown to be applicable to earth-rotation-synthesis observations in which some of the sources vary significantly in brightness over the course of the observation. Sources which vary over the course of a day are not very common, but they do occur, and are sometimes (in the case for example of novae) sources which it is of the highest interest to map accurately. But in fact one does not have to look for natural variations in flux density to encounter this problem: any movement of the primary beam of the array on the sky during an observation will generate artificial fluctuations, not only in the average flux density of sources, but, due to the frequency-dependent size of the primary beam, also in their spectral indices. Since some kind of pointing error in a mechanically tracking dish is probably unavoidable, this is likely to be a problematic issue when performing any kind of wide-field or mosaiced observation (see e.g. Bhatnagar et al \cite{bhatnagar}), particularly so in view of current hopes for improvements in dynamic range from the several wide-band, dish-antenna arrays, such as eVLA, eMERLIN and MeerKAT, which are currently under construction.

Deconvolution of time-varying sources reveals some issues which are not usually encountered in the frequency-synthesis case. The principal one of these is that choice of basis function is now of some importance. This issue was explored with some care, in particular the avoidance of Gibbs ringing when periodic basis functions are employed. It was also shown that, provided the time variation in the field is limited to a single, unresolved source, a particularly simple 2-beam technique can produce an almost perfect deconvolution.

Whereas the original parallel decomposition treatment employed at most 3 or 4 beams, some of the situations described in the present paper required as many as 30. Use of such large numbers of beams gives rise to a computational difficulty due to the requirement in the original Sault-Wieringa algorithm to store, for $N$ input beams, of order $N^2$ cross-correlation images. It is shown here that this memory load can be much reduced if the set of beams is made orthogonal. An algorithm for doing this was proposed and it was shown how `de-orthogonalized' clean components can be recovered at the end of the cleaning process via a simple matrix inversion.

\begin{acknowledgements}
We thank the anonymous referee for some sound advice and for directing our attention to the compressive sensing papers.
\end{acknowledgements}

\end{document}